# Alternate islands of multiple isochronous chains in wave-particle interactions

M. C. de Sousa[1,*], I. L. Caldas[1], A. M. Ozorio de Almeida[2], F. B. Rizzato[3], R. Pakter[3]

[1] *Instituto de Física, Universidade de São Paulo, São Paulo, Brazil*
[2] *Centro Brasileiro de Pesquisas Físicas, Rio de Janeiro, Brazil*
[3] *Instituto de Física, Universidade Federal do Rio Grande do Sul, Porto Alegre, Brazil*

We analyze the dynamics of a relativistic particle moving in a uniform magnetic field and perturbed by a standing electrostatic wave. We show that a pulsed wave produces an infinite number of perturbative terms with the same winding number, which may generate islands in the same region of phase space. As a consequence, the number of isochronous island chains varies as a function of the wave parameters. We observe that in all the resonances, the number of chains is related to the amplitude of the various resonant terms. We determine analytically the position of the periodic points and the number of island chains as a function of the wave number and wave period. Such information is very important when one is concerned with regular particle acceleration, since it is necessary to adjust the initial conditions of the particle to obtain the maximum acceleration.

Wave-particle interaction is basically a nonlinear process [1, 2] that may present regular and chaotic trajectories in its phase space [3]. This kind of interaction can be found in many areas of physics [1, 4-6], and it is used in a wide range of applications as an efficient way for particle heating [1, 7-9] and particle acceleration [1, 5, 9-11].

References [12-14] present two cases of particles moving in a uniform magnetic field and perturbed by electrostatic waves. For such systems, appropriate resonant conditions are responsible for a great amount of particle acceleration. References [12, 13] determine the parameters values for which the acceleration may be maximum.

However, the process of regular acceleration also depends on the trajectory followed by the particles. To attain the condition of maximum acceleration, it is necessary to know the position of the resonances and the number of island chains as a function of the parameters. In this way, it is possible to properly adjust the initial conditions of the particles and make them follow the best trajectory in phase space. Moreover, successive bifurcations changing the number of chains modify the acceleration conditions. Even so, these bifurcations have not been explored yet.

In this report, we analyze the onset of different island chains described by the dynamics of a twist system consisting of a relativistic particle moving in a uniform magnetic field. This integrable system is kicked by standing electrostatic pulses [15-17], such that it becomes near-integrable for small amplitudes [1-3, 12, 18].

Expanding the pulses in a Fourier-Bessel series, we observe that the wave presents an infinite number of perturbative terms. There are groups of perturbative terms with the same winding number that may generate different islands in the same region of phase space. This superposition alters the number of island chains according to the wave parameters. Using the map of the system, we carry a series of analytical estimates, including the position of the periodic points in phase space and the number of island chains as functions of the wave parameters.

The Hamiltonian of the system also reveals that the perturbative terms are symmetric, and as a consequence, the total number of islands is even for all the resonances. Thus, for any resonance, the number of chains is even when the number of islands in each chain is odd. On the other hand, when the number of islands in each chain is even, the number of chains may be even or odd.

Following Refs. [13, 14], we analyze a beam of charged particles interacting with a uniform magnetic field and a standing electrostatic wave. We assume that the density of the beam is very low, so that it does not induce any wave growth, and the particles may be considered as test particles that do not interact with each other. The dimensionless Hamiltonian that describes the dynamics transverse to the magnetic field is given by [13, 14]

$$H(I, \theta, t) = \sqrt{1+2I} + \frac{\varepsilon}{2} \cos(k\sqrt{2I} \sin\theta) \sum_{n=-\infty}^{+\infty} \delta(t - nT), \quad (1)$$

where the periodic sum of delta function kicks describes a pulsed wave with dimensionless wave number $k$, period $T$ and amplitude $\varepsilon/2$.

Integrating the system, we obtain an exact map that describes its time evolution [13, 14]:

$$\begin{aligned}
I_{n+1} &= \frac{1}{2}\left\{ 2I_n \sin^2\theta_n + \left[ \sqrt{2I_n}\cos\theta_n + \frac{1}{2}\varepsilon k \sin(k\sqrt{2I_n}\sin\theta_n) \right]^2 \right\}, \\
\theta_{n+1} &= \arctan\left[ \frac{2\sqrt{2I_n}\sin\theta_n}{2\sqrt{2I_n}\cos\theta_n + \varepsilon k \sin(k\sqrt{2I_n}\sin\theta_n)} \right] \\
&\quad + \frac{T}{\sqrt{1+2I_{n+1}}} \pmod{2\pi},
\end{aligned} \quad (2)$$

where $(I_n, \theta_n)$ are defined as the values of $(I, \theta)$ immediately before kick $n$.

This model does not include any effect regarding the radiation emitted by an accelerated charged particle. Typically, particle accelerators present magnetic fields on the order of 1 Tesla. Considering an electron with velocity $v_\perp \cong 0.9999\, c$ perpendicular to the magnetic field, the energy emitted by the electron in a cyclotron period is approximately $4.87 \times 10^{-10} K$ [19, 20], where $K$ denotes its kinetic energy. On the other hand, the maximum energy emitted by the electron due to one wave pulse is estimated as $2.21 \times 10^{-8} K$ [19, 20], for an ultrashort pulse with time duration on the order of $10^{-16}$ seconds [21], $k = 10$ and $\varepsilon = 1$. Therefore, the energy emitted by the electron is much smaller than its kinetic energy, and the radiation process, for our purposes, can be neglected.

* meirielenso@gmail.com

The position of primary resonances in phase space can be estimated if we expand Hamiltonian (1) and map (2) to first order. Expanding Hamiltonian (1) in a Fourier-Bessel series, we obtain

$$H = \sqrt{1+2I} + \frac{\varepsilon}{2T} \sum_{s=-\infty}^{+\infty} \sum_{r=-\infty}^{+\infty} J_r(k\sqrt{2I}) \cos\left(r\theta - \frac{2\pi st}{T}\right). \quad (3)$$

From (3), we calculate the approximate resonant condition

$$\frac{d}{dt}\left(r\theta - \frac{2\pi st}{T}\right) = 0, \quad r\frac{d\theta}{dt} = \frac{2\pi s}{T}, \quad r\omega_0 \cong s\omega, \quad (4)$$

where $\omega = 2\pi/T$ is the frequency of the wave and we approximate $d\theta/dt \cong \omega_0$, with $\omega_0$, the unperturbed frequency, given by

$$\omega_0 = \left.\frac{d\theta}{dt}\right|_{H=H_0} = \frac{dH_0}{dI} = \frac{1}{\sqrt{1+2I}}. \quad (5)$$

Since $\omega$ and $\omega_0$ are both positive, (4) is satisfied only when $r$ and $s$ have the same sign (with $r \neq 0$ and $s \neq 0$). Condition (4) implies that the system behaves resonantly whenever the ratio $\Omega = \omega_0/\omega$ is a rational number [3, 22-28]. The quantity $\Omega$ is called winding number or rotation number, and for $\Omega = s/r$, the periodic points of the resonance repeat themselves after $r$ iterations of the map.

Inserting (5) into (4), we obtain the position of the $(r,s)$ primary resonances with respect to the action variable as

$$I_{r,s} \cong \frac{1}{2}\left(\frac{r}{s\omega}\right)^2 - \frac{1}{2} = \frac{1}{8}\left(\frac{rT}{s\pi}\right)^2 - \frac{1}{2}. \quad (6)$$

To estimate the values of $\theta$ in the periodic points, we approximate the first equation of map (2) to first order in $\varepsilon$ such that

$$I_{n+1} = I_n + \frac{\varepsilon k}{2}\sqrt{2I_n} \cos\theta_n \sin(k\sqrt{2I_n} \sin\theta_n). \quad (7)$$

Following this first order approximation in the successive iterations of the map, all the terms on the order of $O(\varepsilon^2)$ should be neglected, and thus we have

$$I_{n+2} = I_n + \frac{\varepsilon k}{2}\sqrt{2I_n} \cos\theta_n \sin(k\sqrt{2I_n} \sin\theta_n)$$
$$+ \frac{\varepsilon k}{2}\sqrt{2I_n} \cos\theta_{n+1} \sin(k\sqrt{2I_n} \sin\theta_{n+1}),$$

$$I_{n+r} = I_n + \frac{\varepsilon k}{2}\sum_{j=0}^{r-1}\sqrt{2I_n} \cos\theta_{n+j} \sin(k\sqrt{2I_n} \sin\theta_{n+j}), \quad (8)$$

such that the sum in (8) contains just terms of zero order in $\varepsilon$.

As the periodic points return after $r$ iterations of the map, we obtain $I_{n+r} = I_n \cong I_{r,s}$. This result must be valid for $k \neq 0$, $I_{r,s} \neq 0$ and small finite $\varepsilon$. From (8) it follows that

$$G_{r,s} \equiv \sum_{j=0}^{r-1} \cos\theta_{n+j} \sin(k\sqrt{2I_{r,s}} \sin\theta_{n+j}) = 0. \quad (9)$$

Moreover, for the unperturbed periodic points $\theta_{n+1} = \theta_n + 2\pi s/r$, so that $\theta_{n+r} = \theta_n + 2\pi s = \theta_n \pmod{2\pi}$:

$$G_{r,s}(\theta) = \sum_{j=0}^{r-1}\left\{\cos\left(\theta + \frac{2\pi s}{r}j\right)\right.$$
$$\left.\times \sin\left[k\sqrt{2I_{r,s}} \sin\left(\theta + \frac{2\pi s}{r}j\right)\right]\right\} = 0. \quad (10)$$

According to the Poincaré-Birkhoff Fixed Point Theorem, any $(r,s)$ primary resonance presents $Mr$ ($M$ a positive integer) resonant islands around the stable periodic points [3, 27-29]. The set of islands surrounding one stable orbit form a chain [3, 24-29]. In this sense, $M$ indicates the number of isochronous chains and $r$ represents the number of islands in each chain.

Although the Poincaré-Birkhoff Fixed Point Theorem does not determine the value of $M$ [3, 27-29], we generally find in the literature twist systems that present just one chain with $r$ islands [3]. However, for some twist systems more than one chain has already been observed [30, 31]. In our system, more chains are commonly present in the phase space, as it can be seen in Fig. 1.

Figure 1 shows the phase space of the system for $T = 2.5\pi$, $k = 5$ and $\varepsilon = 0.006$. In this figure, there are 6 resonant islands at $I_{6,5} \cong 0.62$. All these islands form a single chain, because just one initial condition visits the 6 islands. In contrast, we need 2 different initial conditions to create all the islands of the (4,3) resonance located at $I_{4,3} \cong 0.89$. Thus, the (4,3) resonance presents 2 chains with 4 islands each. The islands of both chains alternate in the phase space. The different chains are drawn in red and green in Fig. 1, and they can also be distinguished by their different size and shape.

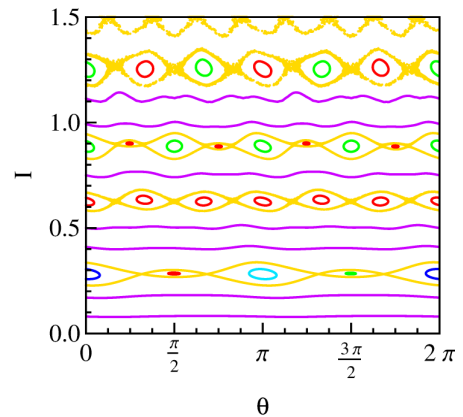

FIG. 1 (Color online). Phase space of the system for $T = 2.5\pi$, $k = 5$ and $\varepsilon = 0.006$. The resonant islands are drawn in red, green, blue and cyan. Each color represents a different island chain. In this figure, we show the (1,1), (6,5), (4,3) and (3,2) resonances located respectively at $I_{1,1} \cong 0.28$, $I_{6,5} \cong 0.62$, $I_{4,3} \cong 0.89$ and $I_{3,2} \cong 1.26$.

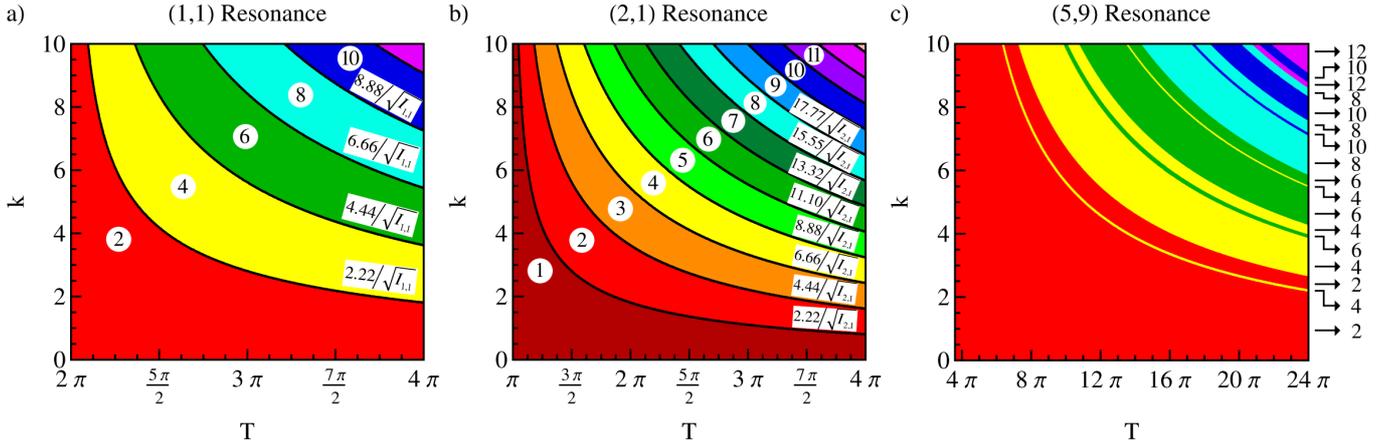

FIG. 2 (Color online). Number of island chains as a function of $T$ and $k$ for the a) (1,1); b) (2,1) and c) (5,9) resonances. Each color represents a different number of chains as indicated in the pictures. Panels (a) and (b) display the expressions obtained numerically for $k$ as a function of $I_{r,s}(T)$ for the black curves separating two regions with a different number of chains.

For the present system, appropriate resonant conditions are useful for particle acceleration [13, 14]. To attain the maximum acceleration, it is necessary to adjust the parameters of the system in such a way that the particle receives a great amount of energy from the wave [13]. Moreover, the initial conditions of the particle should be chosen according to the resonance that presents the best acceleration condition. In this sense, it is important to know the position and the number of island chains for the $(r,s)$ resonances and how this number varies as a function of the wave parameters.

Every $(r,s)$ primary resonance lies on a rational surface $\Omega = s/r$ [3, 22-28]. However, Hamiltonian (3) presents an infinite number of perturbative terms with the same winding number. From this point on, we will always refer to $r$ and $s$ as two positive relative primes. In this case, we observe that all the $(mr, ms)$ perturbations (with $m$ integer and $m \neq 0$) lie on the same rational surface $\Omega = s/r$. Therefore, each of these perturbations may generate isochronous islands in the same region of phase space.

For a twist system that presents only one $(mr, ms)$ perturbative term, $|m|$ represents the number of chains, and $r$ is the number of islands in each chain. The number of chains in this case is constant irrespective of the values of the parameters. For the analyzed system, there are infinite $(mr, ms)$ perturbations with the same winding number and their coupling makes the number of chains vary according to the wave parameters. Close examination of expressions (4) to (10) shows that they are not functions of the individual numbers $r$ and $s$, but rather of the ratio $\Omega = s/r$. Thus, when we choose the value of $\Omega$ in these expressions, we are actually taking all the $(mr, ms)$ perturbative terms with $\Omega = s/r$, and it is impossible to distinguish each individual term.

However, just some of the $(mr, ms)$ perturbations actually generate islands in phase space. In many cases, the $M$ chains are formed due to the superposition of the $(mr, ms)$ perturbations for which $|m| \leq M$. Although there are exceptions, the amplitude of the perturbative terms for which $|m| > M$ are generally much smaller and these terms do not contribute to the appearance of islands, as we show later.

Figure 2 presents the number of island chains as a function of the wave period $T$ and wave number $k$ for the (1,1), (2,1) and (5,9) resonances. The figure was generated using the estimates from expressions (6) and (10). Overall, in the panels the number of chains increases with $T$ and $k$.

To generate Fig. 2(b), we introduced $\Omega = s/r = 1/2$ in expressions (6) and (10). But as we have just seen, $\Omega = 1/2$ represents all the $(2m, m)$ perturbative terms in Hamiltonian (3). Considering a phase space region close to the (2,1) resonance, we can reduce the Hamiltonian (3) to a local Hamiltonian that presents only the (2,1) resonance and, therefore, is valid only in this region:

$$H_{2,1} = \sqrt{1+2I} + \frac{\varepsilon}{2T} \sum_{\substack{r=-\infty \\ r \text{ even}}}^{+\infty} J_r(k\sqrt{2I}) \cos\left(r\theta - \frac{r\pi t}{T}\right), \quad (11)$$

where $I$ should be close to $I_{2,1}$.

From Hamiltonian (11), we observe that the amplitude of the perturbative terms generating the (2,1) resonance depends on $r$ as $J_r(k\sqrt{(T/\pi)^2 - 1})$. For $k \to 0$ or $T \to \pi$, the argument of the Bessel functions $J_r$ tends to zero. In this case, the amplitude of $J_2$ is much higher than the amplitude of the others $J_r$ for which $|r| > 2$. Thus, only the perturbative terms $r = \pm 2$ are relevant, and the (2,1) resonance presents just one chain.

Increasing the values of $T$ or $k$, the amplitude of $J_4$ cannot be neglected anymore. In this case, we have four resonant terms ($r = \pm 2$ and $r = \pm 4$) with considerable amplitude and the phase space presents two chains. As we continue increasing the values of $T$ or $k$, more perturbative terms should be taken into account and the number of island chains increases as shown in Fig. 2(b).

A similar analysis can be carried for the (1,1) resonance shown in Fig. 2(a). However, for the $(r,s)$ resonances with

$r > 4$, the coupling of the resonant terms becomes more complicated and the number of chains does not increase monotonically with $T$ and $k$, as it can be seen in Fig. 2(c) for the (5,9) resonance. In this figure, the regions corresponding to different numbers of chains are altered and as we increase the values of $T$ and $k$, the pattern becomes more complicated.

From Fig. 2, we observe that the number of chains for the (2,1) resonance may be even or odd, but the number of chains for the (1,1) and (5,9) resonances is always even. Considering just the resonant terms of the Hamiltonian (3), we observe that they are symmetric and all the terms for which $r$ is odd are canceled in pairs. Just the terms with even $r$ act on the system, and they generate an even number of islands for every $(r,s)$ primary resonance. Thus, the resonances with odd $r$ always present an even number of chains, whereas for even $r$, the number of chains may be even or odd.

Figures 2(a) and 2(c) show that the resonances with odd $r$ never present a single chain. Even for small values of the wave parameters, these resonances exhibit multiple isochronous chains in phase space. Observing the panels in Fig. 2, we also expect that for sufficiently large values of $T$ and $k$, all the resonances present more than one chain in phase space.

Using condition (10), it is possible to calculate analytically the parameter region for a given number of chains in the $(1,s)$ and $(2,s)$ resonances. Replacing $r = 1$ (or $r = 2$) in condition (10), we obtain the total number of periodic points as a function of $k$ and $I_{1,s}$ (or $I_{2,s}$). Since the resonances present $Mr$ islands [3, 27-29], the parameter region for which the $(1,s)$ resonances present $M$ island chains is given by

$$\frac{(M-2)\pi}{2\sqrt{2I_{1,s}}} < k < \frac{M\pi}{2\sqrt{2I_{1,s}}}, \qquad (12)$$

whereas for $r = 2$, the parameter region is given by

$$\frac{(M-1)\pi}{\sqrt{2I_{2,s}}} < k < \frac{M\pi}{\sqrt{2I_{2,s}}}. \qquad (13)$$

The extremes of expressions (12) and (13) (for $s = 1$) correspond to the black curves in Figs. 2(a) and 2(b). Each black curve separates two regions characterized by a different number of chains. The analytical estimates (12) and (13) agree quite well with the numerical results displayed in these figures.

For the $(r,s)$ resonances with $r > 2$, condition (10) becomes more complicated and it is not possible to calculate analytically the curves separating two parameter regions with a different number of island chains. However, fitting these curves numerically, we find that all of them follow the same kind of power law

$$F(T) \equiv \left[\frac{1}{4}\left(\frac{rT}{s\pi}\right)^2 - 1\right]^{-1/2} = \frac{1}{\sqrt{2I_{r,s}}}. \qquad (14)$$

Expression (14) is valid for all $(r,s)$ resonances, including the resonances that present a more complicated behavior, such as the (5,9) resonance in Fig. 2(c). Thus, the decay of the above-mentioned curves is inversely proportional to the argument of the Bessel functions $J_r(k\sqrt{2I_{r,s}})$ that determine the amplitude of the perturbative terms in (3).

The dynamics of a relativistic particle moving in a uniform magnetic field, perturbed by standing electrostatic kicks, presents an infinite number of perturbative terms with the same winding number, which may generate islands in the same region of phase space. This superposition alters the number of island chains as a function of the wave parameters. When the wave number or wave period are close to their minimum values, the number of chains in phase space is also minimum. As we increase the wave number or wave period, the number of chains increases as well.

Since the islands of the system can be used for regular acceleration [12-14], the variation in the number of chains according to the wave parameters is an important phenomenon. To accelerate the particle in the islands of a given resonance, it is necessary to know the position of the periodic points and the number of island chains. In this way, the initial conditions of the particle may be adjusted in order to attain the maximum acceleration.

We determined the position of the primary resonances in phase space, and we obtained the range of the wave parameters that corresponds to a given number of island chains for the main resonances. We built the parameter space representing the number of chains as a function of the wave period and wave number.

We also observed that the resonant terms acting on the system are symmetric and, as a consequence, most of the primary resonances present an even number of island chains in the phase space. Thus, we conclude that according to the symmetry of the resonant terms, the dynamics of a near-integrable twist system may be dominated by multiple isochronous chains, instead of the usual scenario of single chains.

For the system analyzed in this report, it is found that the number of chains increases without limit as the parameters of the system increase. We point out that this is not a particular result. This phenomenon occurs in near-integrable twist systems that present an infinite number of resonant terms acting on the same rational surface. However, when the number of resonant terms with the same winding number is finite, the number of chains also varies according to the parameters of the system, but there is a finite set for the possible numbers of chains. This set is determined by the resonant terms that have the same winding number and by their superposition.

**ACKNOWLEDGMENTS**

We acknowledge financial support from São Paulo Research Foundation (FAPESP, Brazil) under Grant Nos. 2011/20794-6 and 2011/19296-1, CNPq (Brazil), CAPES (Brazil) and US-AFOSR under Grant No. FA9550-12-1-0438.